\documentclass[11pt]{article}
\usepackage[dvips]{graphicx}
\usepackage{mathrsfs}
\usepackage{amssymb}
\usepackage{amsmath}
\usepackage{latexsym}
\usepackage{multirow}
\usepackage{setspace}
\title{Mission-Aware Medium Access Control\\in Random Access Networks}
\author{Jaeok Park\footnote{Department of Economics, University of
California, Los Angeles (UCLA), Los Angeles, CA 90095-1477, USA
(e-mail: jpark31@ucla.edu)} and Mihaela van der
Schaar\footnote{Department of Electrical Engineering, University of
California, Los Angeles (UCLA), Los Angeles, CA 90095-1594, USA
(e-mail: mihaela@ee.ucla.edu)} }
\date{}

\begin{document}

\maketitle

\begin{abstract}
We study mission-critical networking in wireless communication
networks, where network users are subject to critical events such as
emergencies and crises. If a critical event occurs to a user, the
user needs to send necessary information for help as early as
possible. However, most existing medium access control (MAC)
protocols are not adequate to meet the urgent need for information
transmission by users in a critical situation. In this paer, we
propose a novel class of MAC protocols that utilize available past
information as well as current information. Our proposed protocols
are mission-aware since they prescribe different transmission
decision rules to users in different situations. We show that the
proposed protocols perform well not only when the system faces a
critical situation but also when there is no critical situation. By
utilizing past information, the proposed protocols coordinate
transmissions by users to achieve high throughput in the normal
phase of operation and to let a user in a critical situation make
successful transmissions while it is in the critical situation.
Moreover, the proposed protocols require short memory and no message
exchanges.


\end{abstract}

\section{Introduction}

Network users may face critical situations where life or livelihood
is at risk. Examples include a fire in a building, a natural
disaster in a region, a heart attack of a patient, and a military
attack by an enemy. When a network user detects a critical event, it
is important for the user to inform relevant rescue parties of the
event as early as possible so that they can take the necessary
measures to mitigate the risk or help affected parties recover. This
paper is concerned about delay in the transmission of information
about critical events in mission-critical networking, which occurs
between the detection of critical events by a network user and the
response to them by a rescue party.


We consider wireless communication networks in which users share a
common channel and contend for access. We approach the problem of
dealing with critical situations from a protocol designer's
perspective at the medium access control (MAC) layer. Since multiple
packets transmitted at the same time result in a collision, MAC
protocols are used to coordinate transmissions by users. Distributed
coordination function (DCF), widely deployed in the IEEE 802.11a/b/g
wireless local area network (WLAN) \cite{ieee}, does not
differentiate users, and thus it is unable to coordinate the
behavior of users in the event of critical situations so that a user
in a critical situation uses the channel while others wait.

The enhanced version of DCF, called enhanced distributed channel
access (EDCA), is deployed in IEEE 802.11e \cite{ieeee} and does
differentiate users according to their access categories. EDCA
specifies different contention window sizes and arbitration
interframe spaces to different access categories, yielding a smaller
medium access delay and more bandwidth for the higher-priority
traffic categories \cite{gu}. However, EDCA is designed to support
applications requiring quality-of-service, and as such it is not
directly applicable to mission-critical networking in wireless
networks. In particular, a user having highest-priority data shares
the channel with other users. Although it obtains higher throughput
than others, EDCA does not allow it to ``capture'' the channel until
it finishes transmitting the highest-priority data.

In this paper, we discuss the problem of the protocol designer
mainly in the context of a slotted Aloha system. The protocol
designer cares about total throughput and fairness in the normal
phase, in which there is no critical situation, while he is
concerned about delay in the critical phase. Also, he takes the
complexity of protocols into consideration in both phases. We show
that the dual objective of the protocol designer --- maximizing
throughput and fairness in the normal phase while minimizing delay
in the critical phase --- can be achieved by a class of MAC
protocols utilizing past information. The proposed protocols have
the following desirable properties:
\begin{enumerate}
\item The system achieves high total throughput while yielding equal
throughput to individual users in the normal phase of operation,
when no user is in a critical situation.
\item Should a critical event occur, the user in a critical situation
captures the channel after a short delay while other users wait
until it transmits all the necessary information.
\item The protocols can be implemented without any message exchange.
In particular, they do not require users to know whether other users
are in a critical situation or not.
\item The protocols are based on short memory, thus requiring only a
small memory space for each user.
\end{enumerate}

Slotted Aloha was first introduced in \cite{rob}. Recently, the
framework of game theory is used to analyze the noncooperative or
cooperative behavior of users in slotted Aloha
\cite{jin}--\cite{ma}. In \cite{jin}, the strategy, or the decision
rule, for a user is simply its transmission probability used over
time to attain its desired throughput. In \cite{mack}, the number of
users contending for the channel varies over time, and users know
the number of users currently in the system. The decision rule for a
user used in \cite{mack} is its transmission probability as a
function of the number of users. Altman et al. \cite{altm2} assume
that information on the number of users in the system is unavailable
to users and that newly arrived packets are always transmitted. The
decision rule in their model is the transmission probability for
backlogged packets. A correlation device is used in \cite{altm}.
With the presence of a correlation device, the decision rule for a
user considered in \cite{altm} is its transmission probability
depending on random signals generated by the correlation device. Ma
et al. \cite{ma} define two states for users, a free state and a
backlogged state, and relax the assumption of \cite{altm2} that
newly arrived packets are always transmitted. The decision rule for
a user in their model is two transmission probabilities used in each
state.

In the game theoretic models above, the strategies are those in
one-shot games even though interactions among users are repeated.
That is, authors consider transmission strategies based only on
\emph{current} information (for example, the number of users,
correlation signals, and the state of packets\footnote{Whether the
current packet is new or backlogged is affected by past outcomes,
but it contains very limited information about the past and can be
considered as the ``label'' of the current packet.}) in contrast to
early work that considers transmission probabilities updated based
on the histories of feedback information on the channel states (for
example, \cite{hajek} and \cite{riv}). We consider strategies as
those in repeated games that depend not only on current information
but also on past information. By opening up this possibility, we can
design a simple distributed protocol that performs well both when
there is a critical event and when there is none.

The rest of the paper is organized as follows. We describe the model
in Section 2 and formulate the problem of the protocol designer in
Section 3. In Section 4, we investigate the various trade-offs that
the protocol designer faces and introduce our mission-aware MAC
protocols. We extend the protocols to more general scenarios in
Section 6. We conclude the paper in Section 7.

\section{Model}

We consider an idealized slotted Aloha system as in \cite{hamed}.
Users (pairs of transmitter-receiver nodes) share a communication
channel though which they transmit packets. The total number of
users is $N$, and the set of users is denoted by $\mathcal{N} = \{
1, \ldots, N \}$. We assume that the number of users is fixed over
time and known to users. In the case that users do not know the
total number of users, they can estimate it by using techniques such
as the one in \cite{bianchi2}, and the MAC protocols in this paper
can be modified by replacing the actual number of users with an
estimate.

Time is slotted, and slots are synchronized. We label slots by $t =
1,2,\ldots$. Packets are of the same size, and each packet requires
one slot for transmission. A user always has a packet to transmit
and makes a decision on whether to transmit or not in every slot
\cite{jin} \cite{ma}. The action space of a user can be written as
$A = \{T,W\}$, where $T$ stands for ``transmit'' and $W$ for
``wait.'' We denote the action of user $i$ by $a_i \in A$ and an
action profile or \emph{outcome} by $\mathbf{a} = (a_1,\ldots,a_N)$.
The set of outcomes is denoted by $\mathcal{A} \triangleq A^N$.

A packet is successfully transmitted if it is the only transmission
in the slot. If there is more than one transmission, a collision
occurs. If the transmission of a packet results in a collision, it
is retransmitted in some later slot until it is successfully
received. We assume that user $i$ senses whether the channel is
\emph{idle} (no transmission) or \emph{busy} (at least one
transmission) when it waits. We also assume that the receiver node
sends an acknowledgement signal to the transmitter node when the
transmission is successful. In this way, a user learns whether its
transmission is successful (\emph{success}) or not (\emph{failure}).
Hence, from a user's point of view, there are four possible channel
states, and we define the set of channel states by $S \triangleq \{
idle, busy, success, failure \}$. We use $s_i \in S$ to denote the
channel state of user $i$.

The system is subject to critical events such as emergencies and
crises. When a critical event occurs, it assigns a user to carry out
a mission of describing it to a rescue party. The amount of
information required to describe the critical event depends on the
nature of the particular event, and we model this feature by
assuming that the number of packets required to complete a mission
is determined by a random variable $X$. $X$ takes a value of a
positive integer, and we use $x$ to denote the realized value of
$X$. We call $x$ the \emph{length} of a mission. We assume that $x$
is known only to the user to whom the mission is assigned. We say
that a user is in a \emph{critical situation} if it has a mission
and in a \emph{normal situation} otherwise. We denote the situation
of user $i$ by $y_i \in Y$ where $Y \triangleq \{normal,
critical\}$. We use $\mathbf{y} \triangleq (y_1,\ldots,y_N) \in
\mathcal{Y} \triangleq Y^N$ to denote the entire situations of the
system.

We say that the system is in the \emph{normal phase} when every user
is in a normal situation and in the \emph{critical phase} when some
user is in a critical situation. We assume that there can be at most
one mission in the system at a time. We find this assumption
realistic, considering typically a small number of users who share a
wireless channel and the low frequency of critical events. Our
mission-aware MAC protocols are developed based on this assumption,
but we relax this assumption later in Section 5 and show that the
protocols can be modified to deal with multiple missions at a time.

A user knows about its own past and current situations as well as
its own past channel states. We define the \emph{history} of user
$i$ in slot $t$ as all information that user $i$ has at the
beginning of slot $t$, which can be written as
\begin{eqnarray} 
H_i^t = (y_i^1, s_i^1; \ldots; y_i^{t-1}, s_i^{t-1}; y_i^t),
\end{eqnarray}
for $t = 1,2,\ldots$. Let $\mathcal{H}_t \triangleq (Y \times
S)^{t-1} \times Y$ be the set of all possible histories for a user
in slot $t$. Then the set of all possible histories can be defined
by $\mathcal{H} \triangleq \cup_{t=1}^{\infty} \mathcal{H}_t$.

A \emph{decision rule} specifies a transmission probability
following each history, and thus it can be represented by a mapping
from $\mathcal{H}$ to $[0,1]$. Let $\mathbb{N}_+ \triangleq
\{0,1,\ldots\}$ be the set of nonnegative integers. If a decision
rule depends only on information obtained in the recent $m$ previous
slots and the current slot, we say that it is \emph{based on
$m$-period memory} where $m \in \mathbb{N}_+$. Let $\mathcal{L}_m
\triangleq (Y \times S)^{m} \times Y$ be the set of all $m$-period
histories. Then a decision rule based on $m$-period memory can be
written as
\begin{eqnarray}
f_t: \mathcal{L}_m \rightarrow [0,1].
\end{eqnarray}
$f_t(L_i^t)$ gives the transmission probability for user $i$ in slot
$t$ when the recent $m$-period history of user $i$ is
\begin{eqnarray}
L_i^t = (y_i^{t-m}, s_i^{t-m}; \ldots; y_i^{t-1}, s_i^{t-1}; y_i^t),
\end{eqnarray}
for $t = 1,2,\ldots$. We set $(y_i^{t'}, s_i^{t'}) = (normal, idle)$
for $t' \leq 0$ as a default. A decision rule based on $m$-period
memory is said to be \emph{stationary} if it is independent of $t$.
Let $\mathcal{F}_m$ be the set of all stationary decision rules
based on $m$-period memory. Then the set of all stationary decision
rules based on finite memory is obtained by $\mathcal{F} \triangleq
\cup_{m=0}^{\infty} \mathcal{F}_m$. Given two nonnegative integers
$m_1$ and $m_2$ with $m_1 > m_2$, we say that $f \in
\mathcal{F}_{m_1}$ is \emph{equivalent} to $g \in \mathcal{F}_{m_2}$
if $f(L) = g(L')$ where $L'$ is obtained by deleting information in
the first $(m_1 - m_2)$ slots of $L$, for all $L \in
\mathcal{L}_{m_1}$. If $f \in \mathcal{F}_{m_1}$ is equivalent to $g
\in \mathcal{F}_{m_2}$, then $f$ can be implemented using only
$m_2$-period memory, and thus it can be rewritten as a decision rule
based on $m_2$-period memory.

We define a \emph{protocol} as a profile of stationary decision
rules based on finite memory $\mathbf{f} \triangleq (f_1, \ldots,
f_N) \in \mathcal{F}^N$. Given a protocol, we can derive four
objects: 1) throughput, 2) short-term fairness, 3) expected average
delay, and 4) complexity. We assume that the protocol designer cares
about 1) and 2) in the normal phase and 3) in the critical phase.
The definitions and the importance of these objects will be
explained in the next section. The protocol designer is concerned
about 4) overall. The complexity of a protocol can be defined as
follows.

Given a protocol $\mathbf{f}$, we first define
\begin{eqnarray}
m_i \triangleq \min \{ m \in \mathbb{N}_+ | \ \exists \ g \in
\mathcal{F}_m \textrm{ such that $f_i$ is equivalent to $g$} \},
\end{eqnarray}
for each $i \in \mathcal{N}$. Then $m_i$ is the minimum length of
memory required to implement the decision rule $f_i$. We take the
maximum of $m_i$ across users to obtain
\begin{eqnarray}
m^*(\mathbf{f}) \triangleq \max \{m_1,\ldots,m_N\}.
\end{eqnarray}
Then $m^*(\mathbf{f})$ is the minimum length of memory required to
implement the protocol $\mathbf{f}$, and we say that the protocol
$\mathbf{f}$ is \emph{based on $m^*$-period memory}. Intuitively, a
protocol is simpler when it is based on shorter memory. Thus, we
call $m^*(\mathbf{f})$ the \emph{complexity level} of the protocol
$\mathbf{f}$.

We assume that one of the objectives of the protocol designer is to
prescribe a protocol with low complexity. In other words, the
protocol designer is inclined to prescribe protocols based on short
memory, for example, one-period memory. Considering the large memory
spaces of computing devices, one may find that using decision rules
based only on one-period memory is too restrictive. However,
one-period memory-based decision rules are easy to follow and robust
to variations on memory and computation constraints. Suppose that
the protocol designer is uncertain about the memory and computation
capacities of individual users. If a failure to follow the
prescribed decision rule by a single user results in a total
breakdown of the system, then the protocol designer wants to provide
a simple protocol to ensure that every user can follow it. Moreover,
analysis with decision rules based on one-period memory is
meaningful in that the performance of protocols based on one-period
memory provides a lower bound on that of more complicated protocols
based on longer memory.

\section{Problem Formulation}

We first consider the problem of the protocol designer separately in
the normal phase and in the critical phase. After discussing the
sub-problems in the two phases, we combine them to formulate the
overall problem of the protocol designer.

\subsection{Problem in the Normal Phase}

We consider a time horizon during which there is no critical
situation. In this case, $y_i^t = normal$ for all $i \in
\mathcal{N}$ and $t$ in the horizon. Since $y_i$ is constant, we can
reduce the domain of a stationary decision rule based on $m$-period
memory from $\mathcal{L}_m$ to $S^m$. If a protocol $\mathbf{f}$ has
a complexity level $m^*$, then a Markov model can be constructed
where the state space of the Markov chain is $\mathcal{A}^{m^*}$. If
$\mathbf{f}$ is chosen so that the induced Markov chain has only one
ergodic class, then there exists a unique stationary distribution
$\pi$ on $\mathcal{A}^{m^*}$ \cite{meyer}. We define the throughput
of user $i$ by
\begin{eqnarray}
\tau_i(\mathbf{f}) \triangleq \sum_{(\mathbf{a}_1, \ldots,
\mathbf{a}_{m^*}) \in \mathcal{A}^{m^*}} \pi(\mathbf{a}_1, \ldots,
\mathbf{a}_{m^*}) \left( \frac{1}{m^*} \sum_{ m = 1}^{m^*}
I(\mathbf{a}_m = \mathbf{a}^i) \right),
\end{eqnarray}
where $m^* = m^*(\mathbf{f})$, $\mathbf{a}^i \in \mathcal{A}$ is the
outcome in which only user $i$ transmits, and $I$ is the indicator
function. That is, the throughput of user $i$ is the frequency of
its success in steady state. The total throughput of the system is
defined by
\begin{eqnarray}
\tau(\mathbf{f}) \triangleq \sum_{i=1}^N \tau_i(\mathbf{f}),
\end{eqnarray}
and the throughput profile by
\begin{eqnarray}
\tau^*(\mathbf{f}) \triangleq (\tau_1(\mathbf{f}), \ldots,
\tau_N(\mathbf{f})).
\end{eqnarray}

The protocol designer can evaluate the throughput profile at least
in two aspects. First, he can measure the utilization of the channel
by total throughput. Hence, considering the efficiency of protocols,
he wants to obtain high total throughput. Second, he may have some
preferences over the distributions of total throughput to users.
This is related to QoS differentiation. In some cases, he may prefer
to treat every user equally. In other cases, he may want to yield
different throughput to different users in a certain proportion.

Given a protocol $\mathbf{f}$, we can compute the expected number of
slots with consecutive successes of user $i$ in steady state. Let
$\theta_i$ be the reciprocal of this expected value. Then $\theta_i
\in [0,1]$, where $\theta_i = 0$ means that the expected value is
infinity. We take the minimum of $\theta_i$ to obtain
\begin{eqnarray}
\theta^*(\mathbf{f}) \triangleq \min \{\theta_1,\ldots,\theta_N\},
\end{eqnarray}
and call $\theta^*(\mathbf{f})$ the \emph{short-term fairness level}
of the protocol $\mathbf{f}$. As $\theta^*$ gets larger, the
expected duration of slots in which the channel is used by one user
becomes shorter. Thus, the protocol designer prefers a protocol with
a high short-term fairness level to guarantee periodic usage of the
channel by users.

Summarizing the discussion so far, the protocol designer's problem
in the normal phase can be formulated as
\begin{eqnarray}
\textrm{(P-Norm)}\quad \max_{\mathbf{f} \in \mathcal{F}^N}
U_N(\tau^*(\mathbf{f}), \theta^*(\mathbf{f}), m^*(\mathbf{f})),
\end{eqnarray}
where $U_N$ is the utility function of the protocol designer in the
normal phase, defined on $[0,1]^N \times [0,1] \times
{\mathbb{N}}_+$. To make the utility function consistent with the
preferences of the protocol designer, we assume that $U_N$ is
increasing in $\tau_i(\mathbf{f})$, for each $i \in \mathcal{N}$,
and $\theta^*(\mathbf{f})$ and decreasing in $m^*(\mathbf{f})$.

\subsection{Problem in the Critical Phase}

Now we consider a time horizon from the start to the end of a
mission. Suppose that a mission is assigned to user $i$ in slot
$t_0$ and that user $i$ completes its mission in slot $t_1$. Then
for $t = t_0, \ldots, t_1$, $y_i^t = critical$ and $y_j^t = normal$
for $j \neq i$. The number of slots needed to complete the mission
is $\hat{x}_i = t_1 - t_0 + 1$. Once a protocol $\mathbf{f} \in
\mathcal{F}^N$ and the $m^*$-period histories of users in slot $t_0$
$\mathbf{L} \triangleq (L_1,\ldots,L_N) \in \mathcal{L}_{m^*}^N$ are
specified, where $m^* = m^*(\mathbf{f})$, we can determine the
probability distribution over the number of slots required for user
$i$ to complete the transmission of $x$ packets. Thus, $\hat{x}_i$
can be considered as a realization of a random variable, called
$\hat{X}_i$, whose probability distribution depends on $x$,
$\mathbf{L}$, and $\mathbf{f}$. We define
$\bar{X}_i(x,\mathbf{L},\mathbf{f})$ as the expected value of
$\hat{X}_i$ given $x$, $\mathbf{L}$, and $\mathbf{f}$. We also
define
\begin{eqnarray}
\bar{D}_i(x,\mathbf{L},\mathbf{f}) \triangleq
\bar{X}_i(x,\mathbf{L},\mathbf{f}) - x,
\end{eqnarray}
which we call the \emph{expected delay} in a mission of user $i$.
$\bar{D}_i(x,\mathbf{L},\mathbf{f})$ represents the expected number
of slots during a critical situation of user $i$ that are not used
for the successful transmission of user $i$'s packets when the
length of the mission is $x$, the $m^*$-period histories of users is
$\mathbf{L}$, and the protocol is $\mathbf{f}$.

$x$ follows the probability distribution of random variable $X$, and
$\mathbf{f}$ induces a stationary distribution on
$\mathcal{L}_{m^*}^N$ using a Markov model. Hence, we can calculate
the expected value of $\bar{D}_i$ given a protocol $\mathbf{f}$ to
obtain
\begin{eqnarray}
D_i(\mathbf{f}) \triangleq
E_{x,\mathbf{L}}[\bar{D}_i(x,\mathbf{L},\mathbf{f})],
\end{eqnarray}
which can be considered as the \emph{average expected delay} in a
mission of user $i$. Finally, we define the \emph{average expected
delay of the protocol $\mathbf{f}$} by
\begin{eqnarray} \label{eq:defdelay}
D^*(\mathbf{f}) \triangleq \max \{D_1(\mathbf{f}), \ldots,
D_N(\mathbf{f}) \}.
\end{eqnarray}
The average expected delay measures the expected number of slots in
which a user with a mission waits or experiences a collision during
its mission. The party affected by a critical event can be rescued
in a timely manner only when the mission is completed without delay.
Therefore, the protocol designer prefers protocols that yield a
small average expected delay. Note that $\hat{x}_i \geq x$ for any
realization of $X$ and $\hat{X}_i$ , and thus $D^*(\mathbf{f}) \geq
0$ for all $\mathbf{f} \in \mathcal{F}^N$.

Suppose that the protocol designer has a utility function in the
critical phase, $U_C$, defined on $[0,+\infty) \times
{\mathbb{N}}_+$. Then the protocol designer's problem in the
critical phase can be formulated as
\begin{eqnarray}
\textrm{(P-Crit)}\quad \max_{\mathbf{f} \in \mathcal{F}^N}
U_C(D^*(\mathbf{f}), m^*(\mathbf{f})),
\end{eqnarray}
where $U_C$ is decreasing in $D^*(\mathbf{f})$ and
$m^*(\mathbf{f})$.

\subsection{Overall Problem}

Depending on the arrival of critical events, the system is in the
critical phase for some slots and in the normal phase for others.
Hence, the protocol designer needs to find a protocol that performs
well in both phases. There may exist a trade-off between the
performance in the normal phase and that in the critical phase. When
facing such a trade-off, the protocol designer needs to find a
protocol that resolves the trade-off by solving the following
overall problem:
\begin{eqnarray}
\textrm{(OP)}\quad &&\max_{\mathbf{f} \in \mathcal{F}^N}
U(\tau^*(\mathbf{f}_N), \theta^*(\mathbf{f}_N), m^*(\mathbf{f}_N),
D^*(\mathbf{f}),m^*(\mathbf{f})),
\end{eqnarray}
where $\mathbf{f}_N$ is the sub-protocol of $\mathbf{f}$ obtained by
fixing $y_i^t = normal$ for all $i$.\footnote{Formally, ${f}_{i,N}$
that constitutes $\mathbf{f}_N$ can be considered as a restriction
of $f_i$ to the subset of $\mathcal{L}_{m_i}$ that contains $y_i =
normal$ only.} $U$ denotes the overall utility function of the
manager, defined on $[0,1]^N \times [0,1] \times {\mathbb{N}}_+
\times [0,+\infty) \times {\mathbb{N}}_+$, and it is increasing in
the first two arguments and decreasing in the last three. In the
formulation, the protocol designer may have different tolerance on
the complexity in the two phases. For example, he may want to keep
complexity low in the normal phase while allowing higher complexity
in the critical phase.

\section{Performance Analysis}

This section investigates various trade-offs between the variables
in the protocol manager's problem. In the normal phase, we analyze
the trade-off between total throughput and short-term fairness by
imposing symmetry and fixing complexity. In the critical phase, we
show the trade-off between the average expected delay and
complexity. Finally, we illustrate the trade-off between short-term
fairness and the average expected delay and between total throughput
and the average expected delay, which are variables of interest in
different phases. The analysis in this section provides results
based on which the protocol designer can choose his optimal protocol
once his utility function is specified.

\subsection{Performance in the Normal Phase}

We analyze the performance in the normal phase using the constrained
optimization approach to (P-Norm). First, we impose a symmetry
constraint which requires every user to follow the same decision
rule. This will be optimal when the protocol manager desires to
yield the same throughput to every user. Second, we fix the
short-term fairness level and the complexity level. By varying the
short-term fairness level and finding optimal values of the
constrained optimization problem, we can trace the trade-off between
total throughput and short-term fairness.

\subsubsection{No Memory}

For tractability, we consider stationary decision rules based on no
memory and one-period memory. We first consider the case where users
do not use past information to determine their transmission
probabilities. In that case, a stationary decision rule is just a
single transmission probability used over time. Imposing the
symmetry constraint, we denote the common transmission probability
by $p$. Then total throughput is given by\footnote{When the protocol
$\mathbf{f}$ prescribes the same decision rule $f$ to every user, we
use $f$ instead of $\mathbf{f}$ as the argument of functions whose
original argument is a protocol.}
\begin{eqnarray}
\tau(p) = N p (1-p)^{N-1},
\end{eqnarray}
and the short-term fairness level is
\begin{eqnarray}
\theta(p) = 1 - p (1-p)^{N-1}.
\end{eqnarray}
Combining these two, we obtain
\begin{eqnarray}
\theta + \frac{\tau}{N} = 1,
\end{eqnarray}
which illustrates a trade-off between total throughput and
short-term fairness. Total throughput is maximized at $p = 1/N$
while the short-term fairness level is maximized at $p = 0$ and 1
where total throughput is zero. Maximum total throughput $(1 -
1/N)^{N-1}$ converges to $1/e \approx 0.368$ as $N \rightarrow
\infty$. Note that this value is equal to the maximum achievable
throughput of the stabilized slotted Aloha system with an infinite
set of nodes \cite{bert}. The short-term fairness level of the
protocol $p = 1/N$ converges to 1 as $N \rightarrow \infty$.

\subsubsection{One-period Memory}

Now we consider stationary decision rules that utilize the channel
states of the previous slot. A stationary decision rule for user $i$
based on one-period memory in the normal phase can be expressed as
$f_i: S \rightarrow [0,1]$. The reciprocal of the expected number of
slots with consecutive successes of user $i$ is given by
\begin{eqnarray}
\theta_i = 1 - f_i(success) \prod_{j \neq i} ( 1 - f_j(busy) ).
\end{eqnarray}
We impose the symmetry constraint on the protocol and use $f$ to
denote the common stationary decision rule based on one-period
memory. By setting the short-term fairness level at $\theta$, we
obtain a constrained version of (P-Norm):
\begin{eqnarray}
\textrm{(P-Norm1)}\quad && \hat{\tau}(\theta) = \max_{f \in
\mathcal{F}_1} \tau(f)\\
&&\qquad \quad \textrm{subject to} \ f(success) ( 1 - f(busy)
)^{N-1} = 1 - \theta. \label{eq:theta}
\end{eqnarray}

We first show that the protocol designer can achieve maximum total
throughput 1 and the maximum short-term fairness level 1 at the same
time with a symmetric stationary decision rule based on one-period
memory when there are only two users.
\newtheorem{prop1}{Proposition}
\begin{prop1}
With $N = 2$, $\hat{\tau}(1) = 1$.
\end{prop1}

\noindent\emph{\textbf{Proof}}: Consider a decision rule $\hat{f}
\in \mathcal{F}_1$ defined by $\hat{f}(idle) = \hat{f}(failure) =
1/2$, $\hat{f}(busy) = 1$, and $\hat{f}(success) = 0$. Note that
$\hat{f}$ satisfies (\ref{eq:theta}) with $\theta = 1$. The
transition probability matrix on $\mathcal{A} = \{ (W,W), (W,T),
(T,W), (T,T) \}$ when both users use $\hat{f}$ is given by
\begin{eqnarray}
P = \left[ \begin{array}{cccc}
\frac{1}{4} & \frac{1}{4} & \frac{1}{4} & \frac{1}{4} \\
0 & 0 & 1 & 0 \\
0 & 1 & 0 & 0 \\
\frac{1}{4} & \frac{1}{4} & \frac{1}{4} & \frac{1}{4}
\end{array} \right].
\end{eqnarray}
From the structure of $P$, we can see that $(W,W)$ and $(T,T)$ are
transient states while $(W,T)$ and $(T,W)$ are ergodic states
\cite{meyer}. Once an ergodic state is reached, $(W,T)$ and $(T,W)$
alternate. Thus, $\tau_1(\hat{f}) = \tau_2(\hat{f}) = 1/2$ and
$\tau(\hat{f}) = 1$. Since $\tau(\mathbf{f}) \leq 1$ for all
$\mathbf{f} \in \mathcal{F}$, $\hat{f}$ attains the maximum of
(P-Norm1). {\hspace{\stretch{1}} \rule{1ex}{1ex}}

\medskip
Proposition 1 shows that channel sharing between two users can be
achieved without communication when they use the decision rule
$\hat{f}$. Initially, they contend with each other with transmission
probability 1/2. Once a user succeeds, they take a turn by
alternating between $T$ and $W$. This perfect channel sharing scheme
is no longer possible with three or more users. If three or more
users use $\hat{f}$, then a success can last only one slot because
it will be followed by a collision for sure, and as a result the
system will be in a collision state most of the time.

Let us partition the set of outcomes $\mathcal{A}$ into $(N+1)$ sets
according to the number of transmissions in outcomes. That is, we
express $\mathcal{A} = \mathcal{A}_0 \cup \cdots \cup \mathcal{A}_N$
where $\mathcal{A}_k$ is the set of outcomes with $k$ transmissions,
for $k = 0,1,\ldots,N$. We obtain an approximate solution to
(P-Norm1) by finding a decision rule $f$ in $\mathcal{F}_1$ that
maximizes \emph{one-step} transition probabilities to
$\mathcal{A}_1$, in which a successful transmission occurs, when
followed by every user.

First, suppose that the outcome in the previous slot is in
$\mathcal{A}_0$, i.e., the channel was idle. Then every user
transmits with probability $f(idle)$. If every user uses the same
transmission probability, say $p$, then the probability of success
is given by $Np(1-p)^{N-1}$, and this expression is maximized at $p
= 1/N$. Hence, we set $f(idle) = 1/N$ to maximize the one-step
transition probability from $\mathcal{A}_0$ to $\mathcal{A}_1$.

Next, suppose that the outcome in the previous slot is in
$\mathcal{A}_1$, i.e., there was a successful transmission. Then one
user transmits with probability $f(success)$ while $(N-1)$ users
with $f(busy)$. The probability of success in the current slot is
given by
\begin{eqnarray} \label{eq:a1a1}
f(success)\left(1-f(busy) \right)^{N-1} +
(N-1)f(busy)\left(1-f(busy) \right)^{N-2}(1 - f(success)).
\end{eqnarray}
The first term in (\ref{eq:a1a1}) is fixed at $1- \theta$ by
(\ref{eq:theta}). The second term is positive if $f(success) < 1$
and $f(busy) > 0$. If $\theta$ is small, however, the second term is
near zero. So we ignore the effect of the second term.

We consider two combinations of $f(success)$ and $f(busy)$ that
satisfy (\ref{eq:theta}):
\begin{eqnarray}
&f(success) = 1 - \theta &\textrm{ and } f(busy) = 0, \quad \textrm{ and} \label{eq:the1}\\
&f(success) = 1 &\textrm{ and } f(busy) = 1 - \sqrt[N-1]{1 -
\theta}. \label{eq:the2}
\end{eqnarray}
Ma et al. \cite{ma} adopt (\ref{eq:the2}) for their two-state
protocol. The main difference between these two combinations is that
with (\ref{eq:the1}) a capture by a user ends when the user releases
the channel whereas with (\ref{eq:the2}) it ends when another user
creates a collision. We choose (\ref{eq:the1}) over (\ref{eq:the2})
for the following two reasons. First, the probabilities in
(\ref{eq:the1}) are independent of the number of users while
$f(busy)$ in (\ref{eq:the2}) depends on it. Thus, (\ref{eq:the1})
will be more robust in achieving a desired duration of consecutive
successes in an environment where the number of users is unknown.
Second, (\ref{eq:the1}) yields a more fair use of the channel than
(\ref{eq:the2}) in the following sense. With (\ref{eq:the1}), when a
capture ends, the channel goes to an idle state in which every user
contends on an equal basis. Hence, a user who captures the channel
next time is chosen equally likely among $N$ users. On the other
hand, since $f(busy) \approx 0$ in (\ref{eq:the2}) when $\theta$ is
not large and $N$ is not small (for example, $f(busy) = 0.0543$ when
$\theta = 0.2$ and $N = 5$), it is most likely that a capture ends
by the transmission of \emph{one} other user. Since $f(busy) \approx
0$, those who waited in the collision are likely to wait until the
contention is resolved between the two users who collided. Hence,
when a capture by a users ends, the same user will capture the
channel again next time with probability near 1/2. This implies that
there are fewer ``changes of hands'' with (\ref{eq:the2}) than with
(\ref{eq:the1}).

Lastly, suppose that the outcome in the previous slot is in
$\mathcal{A}_2$ through $\mathcal{A}_N$, i.e., there was a
collision. The transmission probability that has not been specified
is $f(failure)$. With transmission probabilities chosen so far,
i.e., $f(idle) = 1/N$, $f(busy) = 0$, and $f(success) = 1 - \theta$,
a transition from a success state to a collision state is not
possible, and from an idle state, $\mathcal{A}_2$ is most likely
among $\mathcal{A}_2$ through $\mathcal{A}_N$. Hence, we choose
$f(failure)$ to maximize the one-step transition probability from
$\mathcal{A}_2$ to $\mathcal{A}_1$. Since there are two users who
transmit with $f(failure)$ while others wait following an outcome in
$\mathcal{A}_2$, the one-step transition probability is maximized at
$f(failure) = 1/2$.

The discussion so far provides an approximate solution to the
problem of maximizing one-step transition probabilities to a success
state, which we denote by $\tilde{f}$ where $\tilde{f}(idle) = 1/N$,
$\tilde{f}(busy) = 0$, $\tilde{f}(success) = 1 - \theta$, and
$\tilde{f}(failure) = 1/2$. The next proposition provides a lower
bound on the maximum value of (P-Norm1) by deriving the expression
for $\tau(\tilde{f})$.

\newtheorem{prop2}[prop1]{Proposition}
\begin{prop2}
Suppose $\theta > 0$ in (P-Norm1). Define $q_k = C_{k}^N (1/N)^k
(1-1/N)^{N-k}$ for $k = 0, \ldots, N$. Define recursively from $k =
N$ down to 2 by $J_k(k) = 1$ and
\begin{eqnarray}
J_{k'}(k) = \frac{C_{k}^{k+1}}{2^{k+1} - 1} J_{k'}(k+1) +
\frac{C_{k}^{k+2}}{2^{k+2} - 1} J_{k'}(k+2) + \cdots +
\frac{C_{k}^{k'}}{2^{k'} - 1} J_{k'}(k')
\end{eqnarray}
for $k' = k+1, \ldots, N$. Also, define
\begin{eqnarray}
G_k = \frac{2^k}{2^k - 1} \sum_{j = k}^N J_j(k)q_j
\end{eqnarray}
for $k = 2, \ldots, N$, and
\begin{eqnarray}
G_1(\theta) = \frac{1}{\theta} \left( 1 - q_0 - \sum_{k=2}^N
\frac{G_k}{2^k} \right)
\end{eqnarray}
Then
\begin{eqnarray} \label{eq:lowerb}
\hat{\tau}(\theta) \geq \frac{G_1(\theta)}{1 + G_1(\theta) + G_2 +
\cdots + G_N}.
\end{eqnarray}
If $\theta = 0$, then $\hat{\tau}(0) = 1$.
\end{prop2}

\noindent\emph{\textbf{Proof}}: The lower bound in (\ref{eq:lowerb})
is total throughput attained at $\tilde{f}$. Since every user uses
the same decision rule, we can use
$\{\mathcal{A}_0,\ldots,\mathcal{A}_N\}$ as the set of Markov states
instead of $\mathcal{A}$. Let $P(k'|k)$ be the transition
probability from $\mathcal{A}_k$ to $\mathcal{A}_{k'}$ when
$\tilde{f}$ is used. The transition probabilities are given by
\begin{eqnarray}
P(k'|0) &=& q_{k'} \quad \textrm{for $k'=0,\ldots,N$},\\
P(k'|1) &=& \left\{ \begin{array}{ll}
\theta & \textrm{for $k'=0$}\\
1-\theta & \textrm{for $k'=1$}\\
0 & \textrm{for $k'=2,\ldots,N$},
\end{array} \right.\\
P(k'|k) &=& \left\{ \begin{array}{ll}
\frac{C_{k'}^k}{2^k} & \textrm{for $k'=1,\ldots,k$}\\
0 & \textrm{for $k'=k+1,\ldots,N$}
\end{array} \right. ,\textrm{for $k=2,\ldots,N$}.
\end{eqnarray}
If $\theta = 0$, then $\mathcal{A}_1$ is the unique ergodic state,
and thus $\tau(\tilde{f}) = 1$ implying $\hat{\tau}(0) = 1$. If
$\theta > 0$, then every state of the Markov chain is
positive-recurrent since $P(0|k) > 0$ for all $k = 0,\ldots,N$ and
$P(k'|0) > 0$ for all $k' = 0,\ldots,N$. We denote the unique
stationary distribution by $(\pi_k)_{k=0}^N$ where $\pi_k$ is the
probability of $\mathcal{A}_k$ in steady state. Using the
stationarity condition $\pi_k' = \sum_{k=0}^N P(k'|k) \pi_k$ for $k
= 0,\ldots,N$ (one of them redundant), we obtain $\pi_k = G_k \pi_0$
for $k = 1, \ldots, N$. Imposing the probability condition
$\sum_{k=0}^N \pi_k = 1$, we get
\begin{eqnarray}
\pi_1 = \frac{G_1(\theta)}{1 + G_1(\theta) + G_2 + \cdots + G_N},
\end{eqnarray}
which is total throughput at the approximate solution.
{\hspace{\stretch{1}} \rule{1ex}{1ex}}

\begin{figure}
\begin{center}
\includegraphics[width=0.7\textwidth]{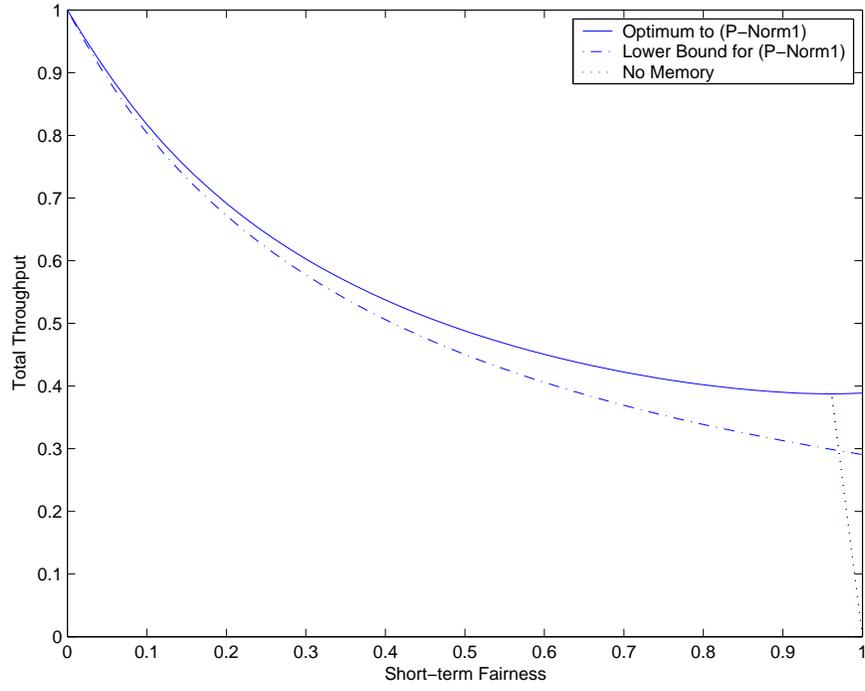}
\caption{Trade-off between throughput and fairness with $N = 10$ }
\label{fig:tradeoff}
\end{center}
\end{figure}

\medskip
$G_2$ through $G_N$ are independent of $\theta$, and $G_1$ is
decreasing in $\theta$. This implies that the lower bound is
decreasing in the short-term fairness level $\theta$, leading to a
trade-off between throughput and fairness. Since $G_1 \rightarrow
\infty$ as $\theta \rightarrow 0$, total throughput can be made
arbitrarily close to 1 by choosing $\theta$ sufficiently small,
which sacrifices fairness. Figure \ref{fig:tradeoff} illustrates the
trade-off between total throughput and the short-term fairness level
at the optimal decision rule to (P-Norm1), which is computed using
numerical methods, and at the approximate solution $\tilde{f}$ with
$N = 10$. Figure \ref{fig:tradeoff} also shows feasible combinations
of throughput and fairness with no memory.

\begin{figure}
\begin{center}
\includegraphics[width=0.7\textwidth]{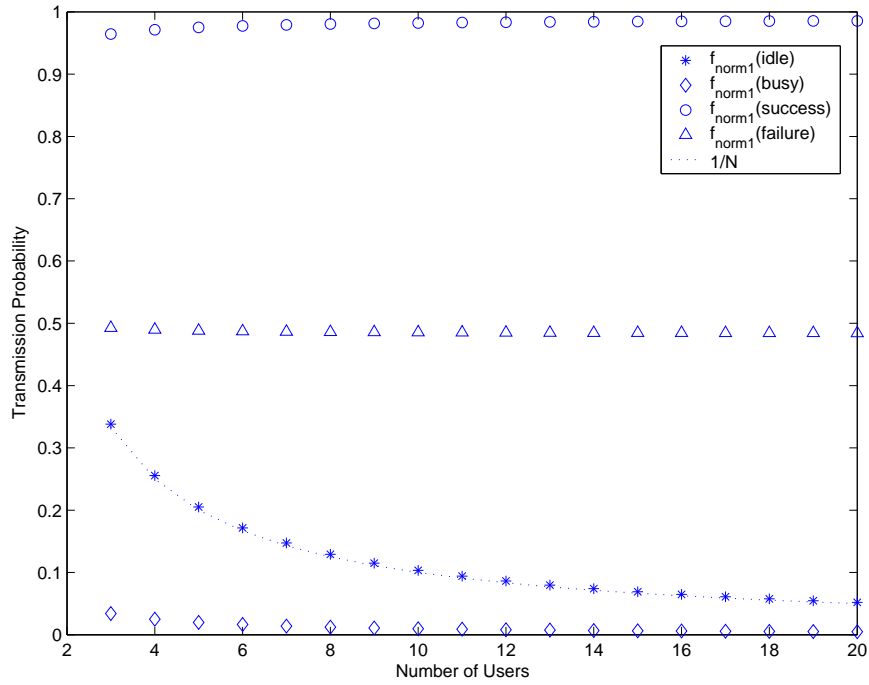}
\caption{Optimal decision rules for (P-Norm1) with $\theta = 0.1$}
\label{fig:optprob}
\end{center}
\end{figure}

\begin{figure}
\begin{center}
\includegraphics[width=0.7\textwidth]{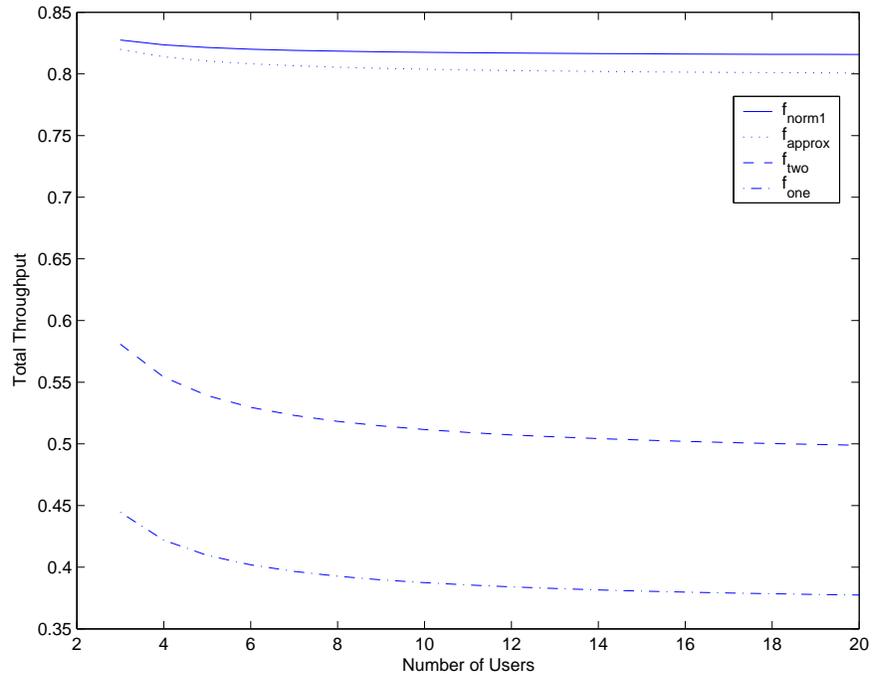}
\caption{Total throughput under different decision rules in the
normal phase ($f_{approx} = \tilde{f}$)} \label{fig:optthrough}
\end{center}
\end{figure}

Let $f_{norm1}$ be the solution to (P-Norm1). We study the structure
of $f_{norm1}$ fixing $\theta = 0.1$ and compare it with
$\tilde{f}$. Again, we rely on numerical methods to compute
$f_{norm1}$. Table \ref{tab:optprob} and Figure \ref{fig:optprob}
show optimal decision rules. $f_{norm1}(idle)$ and
$f_{norm1}(failure)$ are close to those in approximate solution. As
the second term of (\ref{eq:a1a1}) is accounted in the optimal
solution, $f_{norm1}(busy)$ and $f_{norm1}(success)$ take
intermediate values between (\ref{eq:the1}) and (\ref{eq:the2}). We
can see that the approximate solution is quite close to the optimal
solution. As a result, the lower bounds found in Proposition 1 are
close to maximum total throughput as shown in Table
\ref{tab:optthrough} and Figure \ref{fig:optthrough}. Table
\ref{tab:optthrough} and Figure \ref{fig:optthrough} make a
comparison of total throughput under four different decision rules
in the normal phase. A two-state protocol is proposed in \cite{ma}
where users use different transmission probabilities depending on
whether they are in a free state or in a backlogged state. Total
throughput under $\eta$-short-term fairness is given in equation (6)
of \cite{ma}. We set $\eta = 1/\theta = 10$ so that the expected
numbers of slots with consecutive successes are the same under
(\ref{eq:theta}) and under $\eta$-short-term fairness. The total
throughput of the two-state protocol can be obtained by a stationary
decision rule based on one-period memory $f_{two}$ where
$f_{two}(success) = 1$ and $f_{two}(idle) = f_{two}(busy) =
f_{two}(failure) = 1 - \sqrt[N-1]{1 - \frac{1}{\eta}}$. Since
$f_{two}$ does not fully utilize information from the previous slot,
there is a reduction in obtained total throughput compared to that
obtained using $f_{norm1}$. $f_{one} \equiv 1/N$ is the optimal
decision rule based on no memory. Again, utilizing no information
decreases maximum attainable throughput. Note that $f_{norm1}$,
$\tilde{f}$, and $f_{two}$ have the same short-term fairness level
0.1 while that of $f_{one}$ is $1- \frac{1}{N} ( 1 -
\frac{1}{N})^{N-1}$. If users do not use past information, it is not
very likely that a user succeeds for two or more consecutive slots.
As a result, the short-term fairness level of decision rules based
on no memory is very high. For example, $\theta^*(f_{one}) = 0.9613$
when $N = 10$. If we solve (P-Norm1) at $\theta = 1- \frac{1}{N} ( 1
- \frac{1}{N})^{N-1}$, decision rules based on one-period memory
yield no higher total throughput than those based on no memory as
illustrated in Figure \ref{fig:tradeoff}. This implies that the key
feature of decision rules based on one-period memory is their
ability to correlate between successful users in the current slot
and in the future slots. The degree of correlation is determined by
$\theta$. When $\theta$ is close to 1, this correlation does not
exist, and thus utilizing information from the previous slot does
not help to increase throughput.

Finally, we analyze the performance of stationary decision rules
based on one-period memory in an environment of IEEE 802.11 DCF
considered in \cite{bianchi}. Now, the duration of a slot depends on
the state of the channel. Let $\sigma_0$, $\sigma_1$, and $\sigma_2$
be the duration of a slot when the channel state is idle, success,
and collision, respectively. Then total throughput is expressed as
\begin{eqnarray} \label{eq:dcf}
\tau = \frac{P_1 E[P]}{P_0 \sigma_0 + P_1 \sigma_1 + P_2 \sigma_2}
\end{eqnarray}
where $E[P]$ is the average packet payload size and $P_0$, $P_1$,
and $P_2$ are the probabilities of idle, success, and collision
states, respectively. Note that in the idealized slotted Aloha
model, we assume the size of each packet equal to the slot duration
and ignore overhead so that $\sigma_0 = \sigma_1 = \sigma_2 = E[P]$,
and thus the expression for total throughput is reduced to $P_1$,
the probability of success. With stationary decision rules based on
one-period memory, the probabilities can be calculated as $P_0 =
\pi(\mathcal{A}_0)$, $P_1 = \pi(\mathcal{A}_1)$, and $P_2 =
\sum_{k=2}^N \pi(\mathcal{A}_k)$ where $\pi(\mathcal{B})$ is the
probability of outcomes in $\mathcal{B} \subset \mathcal{A}$ in the
stationary distribution.

To obtain numerical results, we use parameters specified by IEEE
802.11a PHY mode-8 \cite{ieeea}, which are tabulated in Table
\ref{tab:spec}. Based on the parameters, we obtain $E[P] = 18432$,
$\sigma_0 = 486$, $\sigma_1 = 22656$, and $\sigma_2 = 21626$ in
bits. We set up a new problem called (P-Norm2) by replacing the
objective function in (P-Norm1) with (\ref{eq:dcf}). We call the
optimal solution to (P-Norm2) $f_{norm2}$. Table \ref{tab:optprob2}
lists the optimal decision rules for (P-Norm2) with $\theta = 0.1$.
Compared to $f_{norm1}$, $f_{norm2}$ prescribes lower transmission
probabilities. Since an idle slot is a lot shorter than a slot in
success or collision states, reaching an idle state is not very
costly compared to reaching a collision state. Hence,
$f_{norm2}(busy)$ and $f_{norm2}(success)$ have the structure of
(\ref{eq:the1}), and $f_{norm2}(idle)$ and $f_{norm2}(failure)$ are
chosen lower than corresponding values in $f_{norm1}$ to avoid
collision states.

\begin{figure}
\begin{center}
\includegraphics[width=0.7\textwidth]{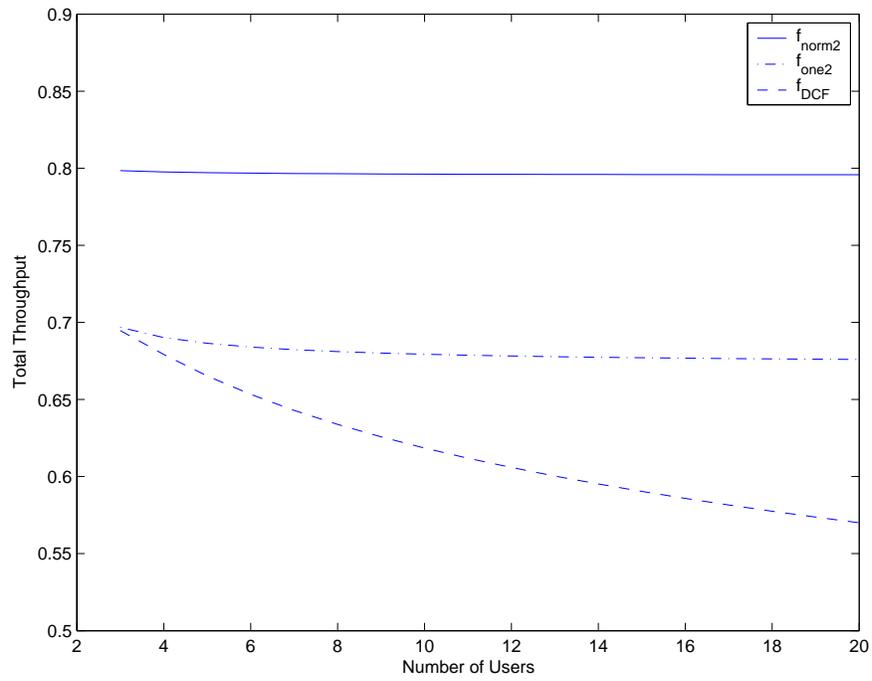}
\caption{Total throughput under different decision rules in the DCF
environment} \label{fig:dcf1}
\end{center}
\end{figure}

Figure \ref{fig:dcf1} compares total throughput in this scenario
under three different decision rules. $f_{one2}$ uses the single
transmission probability that maximizes (\ref{eq:dcf}) whereas
$f_{DCF}$ uses the single transmission probability that corresponds
to the contention window-based exponential backoff (EB) protocol
with $CW_{min} = 16$ and $CW_{max} = 1024$, which can be calculated
using equations (7) and (9) of \cite{bianchi}. We find that the
transmission probabilities derived from DCF are suboptimal as the
number of users increases and that there is a significant
performance improvement by utilizing information obtained in the
previous slot in this environment too.

\begin{figure}
\begin{center}
\includegraphics[width=0.7\textwidth]{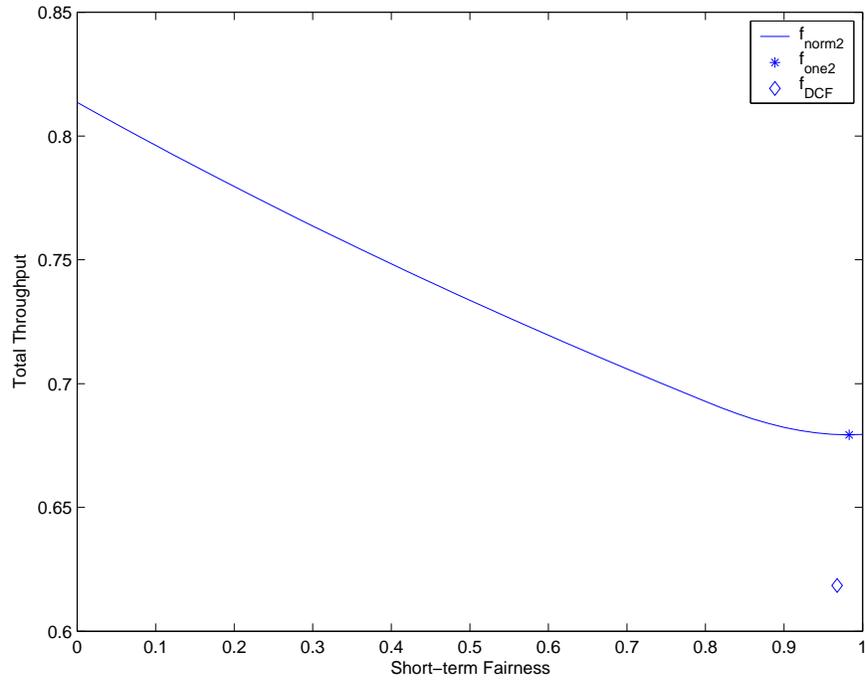}
\caption{Trade-off between throughput and fairness in the DCF
environment with $N = 10$} \label{fig:dcf2}
\end{center}
\end{figure}

Figure \ref{fig:dcf2} illustrates the trade-off between throughput
and fairness in the DCF environment with $N = 10$. As in the slotted
Aloha system, total throughput reduces as the short-term fairness
level increases. The point corresponding to the operation of DCF is
not on the boundary as it operates suboptimally. Again, the gain
from utilizing past information comes from serial correlation among
successful users, which is possible when $\theta$ is not large.
Since there is overhead in DCF, total throughput does not converge
to one as $\theta$ goes to zero.

\subsection{Performance in the Critical Phase}

We now consider slots in which some user is in a critical situation.
As a benchmark case, suppose that the entire situations of the
system is known to all users. Then $y_i^t$ in the histories of user
$i$ is replaced by $\mathbf{y}^t$, and users can adjust their
transmission probabilities depending on others' situations as well
as on their own situations. With the public knowledge of
$\mathbf{y}$, the lower bound for $D^*$ can be attained with a
protocol based on no memory. Define a decision rule $f_0$ in the
critical phase by $p_i^t = 1$ if $y_i^t = critical$, $p_i^t = 0$ if
$y_j^t = critical$ for some $j \neq i$. $f_0$ uses current
information only. Suppose that a mission arrives to user $i$ in slot
$t_0$. If every user follows $f_0$, then user $i$ captures the
channel for $x$ slots starting from slot $t_0$. Then $t_1 =
t_0+x-1$, and we have $\hat{x}_i = x$ for any value of $x$, which
lead to $D^*(f_0) = 0$.

However, the assumption that every user knows the situations of
others is unrealistic considering the distributed nature of wireless
networks. Hence, it is more natural to assume that each user $i$
knows only about its situation, $y_i$. In this scenario, $f_0$
cannot be used since users do not know whether there is another user
who is in a critical situation. Suppose that users use $f_{norm} \in
\mathcal{F}_1$ when they are in a normal situation and $f_{crit}
\equiv 1$ in a critical situation. We impose an important constraint
on $f_{norm}$:
\begin{eqnarray} \label{eq:wait}
f_{norm}(busy) = 0.
\end{eqnarray}
Then the remaining transmission probabilities $f_{norm}(idle)$,
$f_{norm}(success)$, and $f_{norm}(failure)$ determine both total
throughput and the average expected delay while $f_{norm}(success)$
also determines the short-term fairness level by $\theta = 1 -
f_{norm}(success)$ given (\ref{eq:wait}). By varying these three
transmission probabilities, we can obtain the feasible combinations
of total throughput, short-term fairness, and the average expected
delay. In Table \ref{tab:ebprot}, we describe the structure of
$f_{norm}$ and compare it against the persistence probability-based
EB protocol described in \cite{lee}.

Suppose that a mission arrives to user $i$ in slot $t_0$. We examine
the decisions of users using the decision rule that prescribes
$f_{norm}$ in case of a normal situation and $f_{crit}$ in case of a
critical situation, depending on the outcome in slot $t_0 - 1$.
First, we consider the case where user $i$ succeeded in slot $t_0 -
1$, i.e., $\mathbf{a}^{t_0-1} = \mathbf{a}^{i}$. User $i$ transmits
its packet while others wait in slot $t_0$ because user $i$ uses
$f_{crit}$ and user $j \neq i$ uses $f_{norm}$ which prescribes the
transmission probability $p_j^{t_0} = f_{norm}(s_j^{t_0-1}) =
f_{norm}(busy) = 0$ by (\ref{eq:wait}). These decisions remain
unchanged until user $i$ completes its mission in slot $t_0 + x -
1$. When user $i$ switches to $f_{norm}$ in slot $t_0 + x$, it is
expected to capture the channel for additional $1/\theta$ slots. To
prevent this and reset the system, we require that a user in a
critical situation should release the channel when it returns to a
normal situation. The mission-aware protocol described so far is
summarized in Table \ref{tab:prot1} and named as Protocol 1. Note
that Protocol 1 is based on one-period memory. We denote Protocol 1
by $\mathbf{f}_1$. In the case where $\mathbf{L}$ contains
$\mathbf{a}^{i}$ as the most recent outcome,\footnote{$\mathbf{L}$
having $\mathbf{a}$ as the most recent outcome means that $L_i$ has
the channel state for user $i$ in the most recent slot that
corresponds to $\mathbf{a}$.} we have $\hat{x}_i = x$ and thus
\begin{eqnarray} \label{eq:succi}
\bar{D}_i(x,\mathbf{L},\mathbf{f}_1) = 0
\end{eqnarray}
for all $x \in supp(X)$ where $supp(X)$ denotes the support of the
random variable $X$.

Second, we consider the case where some user $j \neq i$ succeeded in
slot $t_0 - 1$, i.e., $\mathbf{a}^{t_0-1} = \mathbf{a}^{j}$. Then
user $i$ transmits in slot $t_0$ because it uses $f_{crit}$, but
user $j$ transmits with probability $1 - \theta$ because
$f_{norm}(success) = 1 - \theta$. Hence, with probability $\theta$
user $i$ starts transmitting its packets from slot $t_0$, and with
probability $1-\theta$ a collision between the packets of user $i$
and $j$ occurs in slot $t_0$. If a collision occurs, then the two
users contend for the channel with $p_i^t = 1$ and $p_j^t =
f_{norm}(failure)$ from slot $t_0+1$ until user $i$ captures the
channel. The number of slots until the first success of user $i$
follows a geometric distribution with parameter $1 -
f_{norm}(failure)$. Hence, we obtain
\begin{eqnarray} \label{eq:succj}
\bar{D}_i(x,\mathbf{L},\mathbf{f}_1) = \frac{1 - \theta}{1 -
f_{norm}(failure)}
\end{eqnarray}
for all $x \in supp(X)$ and $\mathbf{L}$ with $\mathbf{a}^{j}$ as
the most recent outcome. Note, however, that user $j$ learns that
there is a user in a critical situation when encountering a failure
after a success because it cannot happen when every user uses
$f_{norm}$. Again, (\ref{eq:wait}) is crucial for this observation.
Then user $j$ can back off in slot $t_0 + 1$ instead of contending
with the user in a critical situation. This enhancement is
incorporated in Protocol 2 of Table \ref{tab:prot2}, which we denote
by $\mathbf{f}_2$. Note that Protocol 2 is based on two-period
memory. Under Protocol 2, user $i$ starts transmitting in slot $t_0$
with probability $\theta$ and in slot $t_0 + 1$ with probability $1
- \theta$. Therefore, the expected delay is
\begin{eqnarray} \label{eq:succj2}
\bar{D}_i(x,\mathbf{L},\mathbf{f}_2) = 1 - \theta
\end{eqnarray}
for all $x \in supp(X)$ and $\mathbf{L}$ with $\mathbf{a}^{j}$ as
the most recent outcome. Comparing (\ref{eq:succj}) and
(\ref{eq:succj2}), we can see that the higher short-term fairness
level reduces the expected delay for a user if a different user
succeeded in the previous slot. This is true because as $\theta$ is
larger, the probability of yielding gets higher.

Third, we consider the case where a collision occurred in slot $t_0
- 1$. Let $k'$ be the number of users who transmitted in slot $t_0 -
1$ among users other than user $i$. Then according to Protocols 1
and 2, user $i$ transmits with probability 1, $k'$ users transmit
with probability $f_{norm}(failure)$, and the remaining users wait
in slot $t_0$. Note that unlike in the previous case, an inference
about the existence of a critical situation cannot be made because
another collision following a collision is not a zero-probability
event under $f_{norm}$. The collision state will last until user $i$
succeeds. The number of users contending for the channel remains the
same or decreases during collisions, and fixing the number of
contenders at $k'$ will provided an upper bound for the expected
delay. This leads us to
\begin{eqnarray} \label{eq:coll}
\bar{D}_i(x,\mathbf{L},\mathbf{f})  \leq \frac{1}{\left( 1 -
f_{norm}(failure) \right)^{k'}} - 1,
\end{eqnarray}
for all $x \in supp(X)$, for $\mathbf{L}$ such that $k'$ users among
users other than user $i$ transmitted in the most recent outcome,
and for $\mathbf{f} = \mathbf{f}_1, \mathbf{f}_2$. Consider an
outcome with $k$ transmitters, i.e., $\mathbf{a} \in \mathcal{A}_k$.
When users follow the same decision rule, the probability that $k' =
k$, i.e., user $i$ is not one of the $k'$ transmitters, is
$C_k^{N-1}/C_k^N = (N-k)/N$ and the probability that $k' = k - 1$,
i.e., user $i$ is one of the $k$ transmitters, is
$C_{k-1}^{N-1}/C_k^N = k/N$. Hence, we have
\begin{eqnarray} \label{eq:coll2}
\nonumber \bar{D}_i(x,\mathbf{L},\mathbf{f}) &\leq& \frac{N-k}{N}
\cdot \frac{1}{\left( 1 - f_{norm}(failure) \right)^{k}} +
\frac{k}{N}
\cdot \frac{1}{\left( 1 - f_{norm}(failure) \right)^{k-1}} - 1\\
&=& \frac{N - k f_{norm}(failure)}{N \left( 1 - f_{norm}(failure)
\right)^{k}} - 1,
\end{eqnarray}
for all $x \in supp(X)$, $\mathbf{L}$ with $\mathbf{a}^{t_0 - 1} \in
\mathcal{A}_k$, $k = 2, \ldots, N$, and $\mathbf{f} = \mathbf{f}_1,
\mathbf{f}_2$.

Lastly, we consider the case where the channel was idle in slot $t_0
- 1$. Then according to Protocols 1 and 2, user $i$ transmits with
probability 1 while other users transmit with probability
$f_{norm}(idle)$. As in the previous case, no inference about the
existence of a critical situation based on the channel state in slot
$t_0$ can be made because any outcome can be reached following an
idle state under $f_{norm}$. In slot $t_0$, user $i$ succeeds with
probability $(1-f_{norm}(idle))^{N-1}$, and a collision in which one
of transmitters is user $i$ occurs with probability $1 -
(1-f_{norm}(idle))^{N-1}$. Hence, we have
\begin{eqnarray} \label{eq:idle2}
\bar{D}_i(x,\mathbf{L},\mathbf{f}) \leq \sum_{k'=1}^{N-1}
C_{k'}^{N-1} f_{norm}(W,idle)^{k'} (1-f_{norm}(W,idle))^{N-k'-1}
\frac{1}{\left( 1 - f_{norm}(T,failure) \right)^{k'}},
\end{eqnarray}
for all $x \in supp(X)$, $\mathbf{L}$ with $\mathbf{a}^{t_0 - 1} \in
\mathcal{A}_0$, and $\mathbf{f} = \mathbf{f}_1, \mathbf{f}_2$. Since
$f_{norm}$ induces a stationary distribution on $\mathcal{A}$, we
can compute upper bounds on the average expected delays of Protocols
1 and 2 using the definition given in (\ref{eq:defdelay}) and the
results so far.

\newtheorem{prop3}[prop1]{Proposition}
\begin{prop3}
Let $\pi$ be the stationary distribution over $\mathcal{A}$ under
$f_{norm}$. Then for any probability distribution for $X$, the
average expected delays of Protocols 1 and 2 satisfy
\begin{eqnarray} \label{eq:r1}
\nonumber D^*(\mathbf{f}_1) \leq &&\pi(\mathcal{A}_0)
\sum_{k'=1}^{N-1} \frac{ C_{k'}^{N-1} f_{norm}(idle)^{k'}
(1-f_{norm}(idle))^{N-k'-1}}{\left(
1 - f_{norm}(failure) \right)^{k'}} \\
&+& \pi(\mathcal{A}_1) \frac{N-1}{N} \frac{1 - \theta}{1 -
f_{norm}(failure)} + \sum_{k=2}^N \pi(\mathcal{A}_k) \left[ \frac{N
- k f_{norm}(failure)}{N \left( 1 - f_{norm}(failure) \right)^{k}} -
1 \right]
\end{eqnarray}
and
\begin{eqnarray} \label{eq:r2}
\nonumber D^*(\mathbf{f}_2) \leq &&\pi(\mathcal{A}_0)
\sum_{k'=1}^{N-1} \frac{ C_{k'}^{N-1} f_{norm}(idle)^{k'}
(1-f_{norm}(idle))^{N-k'-1}}{\left(
1 - f_{norm}(failure) \right)^{k'}} \\
&+& \pi(\mathcal{A}_1) \frac{N-1}{N} (1 - \theta) + \sum_{k=2}^N
\pi(\mathcal{A}_k) \left[ \frac{N - k f_{norm}(failure)}{N \left( 1
- f_{norm}(failure) \right)^{k}} - 1 \right].
\end{eqnarray}
\end{prop3}

\noindent\emph{\textbf{Proof}}: Since every user uses the same
decision rule under $\mathbf{f}_1$ and $\mathbf{f}_2$,
$D_i(\mathbf{f})$ is the same across users for $\mathbf{f} =
\mathbf{f}_1, \mathbf{f}_2$. Forming a weighted average of
(\ref{eq:succi}), (\ref{eq:succj}), (\ref{eq:coll2}), and
(\ref{eq:idle2}) where the weights are given by
$\pi(\mathcal{A}_1)/N$, $(N-1)\pi(\mathcal{A}_1)/N$,
$\mathcal{A}_k$, and $\mathcal{A}_0$, respectively, we obtain the
upper bound on the average expected delay of Protocol 1 given in
(\ref{eq:r1}). Using (\ref{eq:succj2}) instead of (\ref{eq:succj}),
we obtain the upper bound on the average expected delay of Protocol
2 given in (\ref{eq:r2}). {\hspace{\stretch{1}} \rule{1ex}{1ex}}

\begin{figure}
\begin{center}
\includegraphics[width=0.7\textwidth]{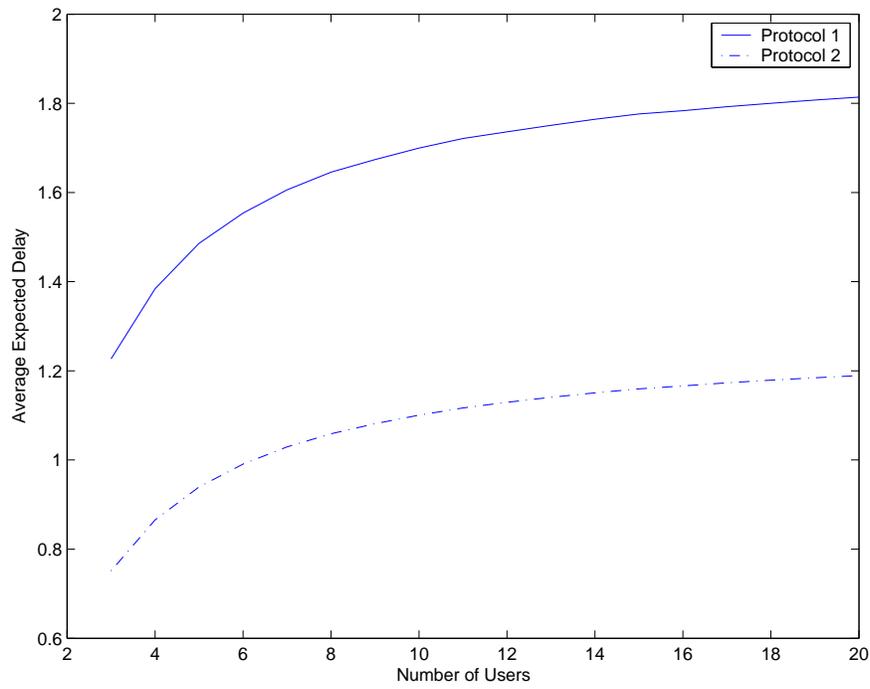}
\caption{Upper bounds on the average expected delays of Protocols 1
and 2} \label{fig:delay}
\end{center}
\end{figure}

\medskip
Figure \ref{fig:delay} plots the upper bounds on the average
expected delays of Protocols 1 and 2 found in Proposition 3 when
$f_{norm}(idle)$ and $f_{norm}(failure)$ are chosen to maximize
total throughput given the constraints $f_{norm}(busy) = 0$ and
$f_{norm}(success) = 0.9$. As the number of users increases, the
average expected delay gets longer. Since a critical event occurs
most likely following a success state ($\pi(\mathcal{A}_1) \approx
0.8$ under $f_{norm}$ with $\theta = 0.1$), the second terms in the
right-hand sides of (\ref{eq:r1}) and (\ref{eq:r2}) dominate the
other terms. As a result, the overestimation used in (\ref{eq:coll})
will not have a large impact on the values of the upper bounds in
Proposition 3, and the upper bounds will be close to the actual
average expected delays. Figure \ref{fig:delay} also shows the
trade-off between the average expected delay and complexity. The
protocol designer can reduce the average expected delay by
increasing the complexity level from 1 to 2.

So far, we have used the average expected delay to measure the
performance of a protocol in the critical phase. Suppose that the
protocol designer is also interested in the worst-case delay as well
as in the average expected delay of a protocol. Both Protocols 1 and
2 have a sequence of outcomes with a positive probability that a
user in a critical situation has to wait for an arbitrary large
number of slots before it starts to transmit, although the
probability of such a sequence of outcomes is close to zero when the
number of waiting slots is large. The protocol designer can bound
realized delays by $m$ with a protocol based on $m$-period memory.
The idea is to make users in a normal situation back off after
experiencing $m$ consecutive collisions so that a user in a critical
situation, if any, can capture the channel. When user $i$ is in a
critical situation, the possible outcomes under Protocols 1 and 2
are either user $i$'s success or a collision. Since the delay can go
infinitely long through consecutive collisions, user $i$ is
guaranteed to start its transmission after $m$ slots at latest if
such modification is applied. Protocol 3 is proposed in Table
\ref{tab:prot3} to introduce this modification. Note that this
modification will have almost no impact on total throughput in the
normal phase because it is very unlikely to have $m$ consecutive
collisions in either phase when $m$ is moderately large, and as a
result it can be thought of as a safety device which is rarely used.

\subsection{Accounting for Both Phases}

We have seen that it is crucial to set $f_{norm}(busy) = 0$ to allow
a user in a critical situation to capture the channel during its
mission without others knowing about the presence of the mission.
The specification of the remaining transmission probabilities
determines the total throughput, the short-term fairness level, and
the average expected delay of Protocols 1 and 2. By varying the
remaining probabilities, the protocol designer can find attainable
combinations of total throughput, short-term fairness, and the
average expected delay, and then he can choose the most preferred
one among them.

\begin{figure}
\begin{center}
\includegraphics[width=0.7\textwidth]{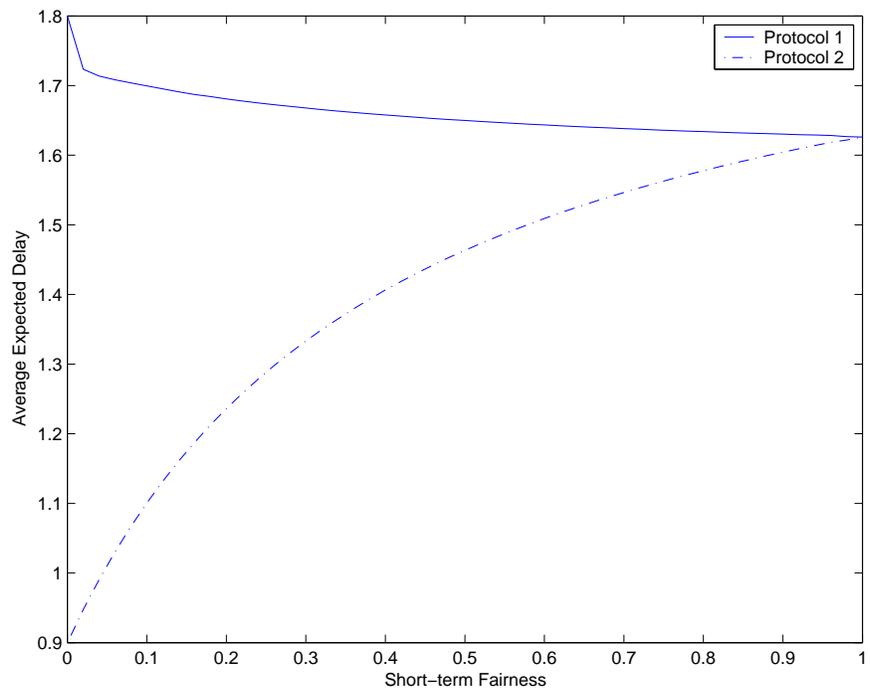}
\caption{Relationship between fairness and delay under Protocols 1
and 2 with $N = 10$} \label{fig:fairdelay}
\end{center}
\end{figure}

We first investigate the relationship between fairness and delay.
Figure \ref{fig:fairdelay} depicts the combinations of short-term
fairness levels and upper bounds on the average expected delay. We
fix $N = 10$ and choose $f_{norm}(idle)$ and $f_{norm}(failure)$ to
maximize total throughput given the constraints $f_{norm}(busy) = 0$
and $f_{norm}(success) = 1 - \theta$. There are two counteracting
effects when the short-term fairness level increases. First, the
system stays in idle and collision states more often as illustrated
in Figure \ref{fig:tradeoff}, and in these states the expected delay
is higher than in success states. Second, the expected delay
decreases when a user other than the one with a mission was
successful in the previous slot, as reflected in the second terms in
the right-hand sides of (\ref{eq:r1}) and (\ref{eq:r2}). The
difference between (\ref{eq:r1}) and (\ref{eq:r2}) is that
$(1-\theta)$ is multiplied by $1/(1 - f_{norm}(failure)) \approx 2$
in (\ref{eq:r1}) while it is not in (\ref{eq:r2}). Thus, the second
effect is stronger in (\ref{eq:r1}) than in (\ref{eq:r2}). Figure
\ref{fig:fairdelay} shows that the second effect is dominant in
(\ref{eq:r1}) while the first in (\ref{eq:r2}). The upper bound on
the average expected delay gets lower as fairness increases with
Protocol 1 whereas it gets higher with Protocol 2.

\begin{figure}
\begin{center}
\includegraphics[width=0.7\textwidth]{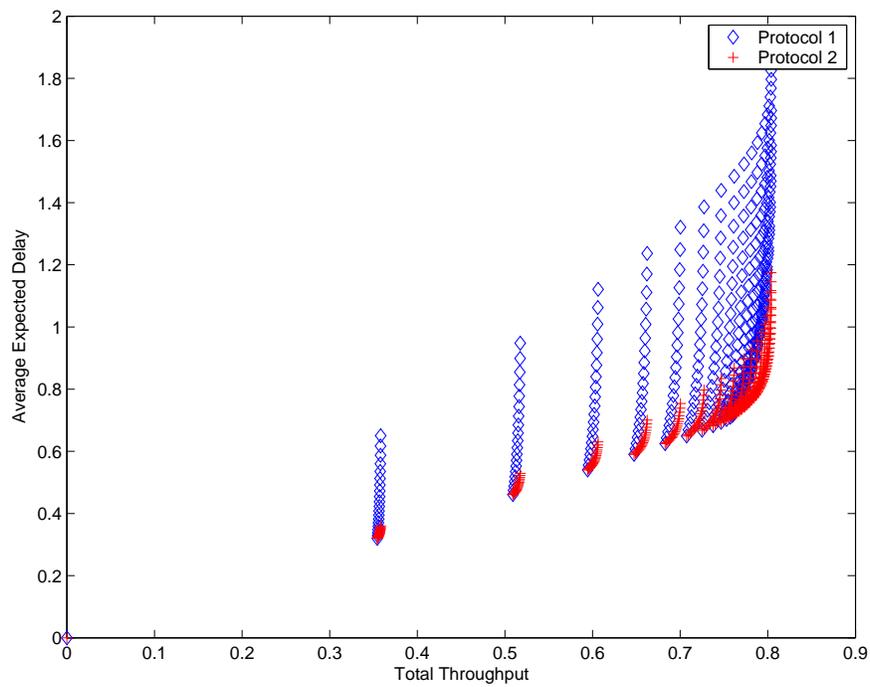}
\caption{Relationship between throughput and delay under Protocols 1
and 2 with $N = 10$} \label{fig:throughdelay}
\end{center}
\end{figure}

Figure \ref{fig:throughdelay} illustrates the trade-off between
throughput and delay with $\theta = 0.1$ and $N = 10$. Given the
transmission probabilities $f_{norm}(idle)$ and $f_{norm}(failure)$
that maximize total throughput fixing $f_{norm}(busy) = 0$ and
$f_{norm}(success) = 1 - \theta$, there is no need to consider
larger transmission probabilities for $f_{norm}(idle)$ and
$f_{norm}(failure)$ because it will decrease total throughput and
increase the average expected delay. Hence, we use values for
$f_{norm}(idle)$ between 0 and 0.11 and for $f_{norm}(failure)$
between 0 and 0.5, and some feasible combinations are shown in
Figure \ref{fig:throughdelay}. The protocol designer can choose the
values for $f_{norm}(idle)$ and $f_{norm}(failure)$ to yield the
most preferred combination of throughput and delay.

\section{Extension to Concurrent Missions}

So far, we have considered a system in which there can be at most
one mission in the system at a time. In this section, we describe
how the proposed protocols can be modified in the presence of
multiple missions.

We first assume that $\mathbf{y}$ is publicly known. Alternatively,
we may assume that $y_i$ is known only to user $i$ but every user
knows the number of missions in the system. We denote $f_{norm1}$
with $N \geq 3$ users by $f_{norm1}(N)$ and define $f_{norm1}(1)
\equiv 1$ and $f_{norm1}(2) \triangleq \hat{f}$ where $\hat{f}$ is
the two-user alternating scheme introduced in Proposition 1. Let
$n^t$ be the number of critical situations in $\mathbf{y}^t$. With
the public knowledge of $\mathbf{y}$, we can consider the following
protocols.
\begin{enumerate}
\item \emph{First-come first-served protocol}\\
Users in a critical situation conduct their missions in the same
order as their missions arrive. That is, if there are users in a
critical situation when a mission arrives to a user, it waits until
all the missions that arrived earlier are completed. If multiple
missions arrive at the same time, the users with these missions
contend with each other with an equal transmission probability to
determine the turn. (Let $n^*$ be the number of missions that
arrived at the same time. Then $n^*$ users transmit with probability
$1/n^*$ until some user succeeds. After the successful user finishes
its mission, the remaining $(n^* - 1)$ users contend with
transmission probability $1/(n^*-1)$ to determine the second user
who uses the channel. This process is repeated until the last user
finishes its mission.)

\item \emph{Sharing protocol}\\
Users in a critical situation use $f_{norm1}(n^t)$ to share the
channel equally while users in a normal situation wait in the
critical phase. Note that there are slots in idle or collision
states when $n^t \geq 3$, which is not the case with the first-come
first-served protocol unless multiple missions arrive at the same
time.
\end{enumerate}

Now we consider the case where each user knows only about its
situation. For the moment, we assume that the system can have at
most two critical situations at a time. We discuss how Protocol 2
can be modified in such a scenario. Suppose that the second mission
arrives to user $j$ in slot $t_0$ while user $i$ is in a critical
situation. Then $\mathbf{a}^{t_0 - 1} = \mathbf{a}^i$, but user $j$
does not know whether the successful user is in a normal situation
or in a critical situation. User $j$ transmits in slot $t_0$. The
transmission by user $j$ informs user $i$ that there exists another
user who is also in a critical situation. If user $i$ were in a
normal situation, it would respond by waiting in slot $t_0 + 1$
according to Protocol 2 so that user $j$ could capture the channel.
However, since user $i$ is in a critical situation, it responds by
transmitting in slot $t_0 + 1$ to inform user $j$ of its critical
situation. Then the presence of two missions becomes a common
knowledge between the two users after two slots. From slot $t_0 + 2$
on, users $i$ and $j$ use $\hat{f}$ to share the channel with the
following modification. In the transient period until one user
succeeds, they always transmit following an idle slot to prevent
other users who are unaware of the missions from taking the channel.
Once one user succeeds, they alternate between $(a_i, a_j) = (T, W)$
and $(a_i, a_j) = (W, T)$ until one of the missions ends. After one
of the missions ends, an idle slot occurs, and the situation becomes
the same as the one with one mission arriving following an idle
slot. We can decrease the expected delay by requiring the user who
completed its mission earlier than the other to wait in the next
slot.

If three or more missions can occur at the same time, then the
dispersion of information on the number of critical situations
through changes in transmission probabilities becomes more
complicated and takes long if possible. Thus, if critical events
occur frequently to multiple users at the same time, the broadcast
by users to signal their critical situations to others will be
valuable in mission-critical networking.

\section{Conclusion}

We have studied the issue of delay in mission-critical networking.
In the context of wireless communication networks, we have proposed
a novel class of MAC protocols that utilize not only current
information but also past information. This allows users to
coordinate their behavior without explicit message exchanges. In the
normal phase, the system can attain high throughput by allowing a
successful user to capture the channel for a period. In the critical
phase, the proposed protocols make a user in a critical situation
capture the channel after a short delay without any message passing
about its critical situation. The proposed protocols fulfill the
objective of the protocol designer in both phases while maintaining
low complexity.

For analytic tractability, we mainly focused on decision rules based
on one-period memory. It will be interesting to investigate the
properties of optimal decision rules based on longer memory such as
two-period memory and how the trade-off between throughput and
fairness changes when longer memory is utilized in the normal phase.
Another potential advantage from utilizing longer memory is the
transmission of more information through the change in transmission
probabilities. One of the reasons that the proposed protocols work
well in a distributed setting is that users can communicate
implicitly through their choices of transmission probabilities. When
the set of possible decision rules expands as longer memory is used,
there are potentially more ``codes'' that can be conveyed through
transmission decisions.

\small
\singlespacing

\newpage
\begin{table}
\caption{Optimal decision rules for (P-Norm1), $f_{norm1}$, with
$\theta = 0.1$} \label{tab:optprob} \centering
\begin{tabular}{|c||c|c|c|c|}
\hline
$N$ & $f_{norm1}(idle)$ & $f_{norm1}(busy)$ & $f_{norm1}(success)$ & $f_{norm1}(failure)$ \\
\hline \hline
3 & $0.338$ & 0.034 & 0.964 & 0.493 \\
\hline
4 & 0.255 & 0.025 & 0.971 & 0.490 \\
\hline
5 & 0.205 & 0.020 & 0.975 & 0.488 \\
\hline
10 & 0.103 & 0.010 & 0.982 & 0.485 \\
\hline
15 & 0.069 & 0.006 & 0.984 & 0.485 \\
\hline
20 & 0.052 & 0.005 & 0.985 & 0.484 \\
\hline
\end{tabular}
\end{table}

\begin{table}
\caption{Comparison of total throughput under different decision
rules in the normal phase} \label{tab:optthrough} \centering
\begin{tabular}{|c||c|c|c|c|}
\hline
$N$ & $f_{norm1}$ & $\tilde{f}$ & $f_{two}$ & $f_{one}$ \\
\hline \hline
3 & 0.8275 & 0.8199 & 0.5808 & 0.4444 \\
\hline
4 & 0.8235 & 0.8139 & 0.5541 & 0.4219 \\
\hline
5 & 0.8214 & 0.8104 & 0.5391 & 0.4096 \\
\hline
10 & 0.8175 & 0.8038 & 0.5116 & 0.3874 \\
\hline
15 & 0.8163 & 0.8017 & 0.5030 & 0.3806 \\
\hline
20 & 0.8157 & 0.8007 & 0.4988 & 0.3774 \\
\hline
\end{tabular}
\end{table}

\begin{table}
\caption{IEEE 802.11a PHY mode-8 parameters} \label{tab:spec}
\centering
\begin{tabular}{|c|c|}
\hline
Parameters & Values \\
\hline \hline
Packet payload & 2304 octets \\
MAC header & 28 octets \\
ACK frame size & 14 octets \\
\hline
Data rate & 54 Mbps \\
Propagation delay & 1 $\mu$s \\
Slot time & 9 $\mu$s \\
PHY header time & 20 $\mu$s \\
SIFS & 16 $\mu$s \\
DIFS & 34 $\mu$s \\
\hline
\end{tabular}
\end{table}

\begin{table}
\caption{Optimal decision rules for (P-Norm2), ${f}_{norm2}$, with
$\theta = 0.1$} \label{tab:optprob2} \centering
\begin{tabular}{|c||c|c|c|c|}
\hline
$N$ & $f_{norm2}(idle)$ & $f_{norm2}(busy)$ & $f_{norm2}(success)$ & $f_{norm2}(failure)$ \\
\hline \hline
3 & $0.077$ & 0 & 0.9 & 0.136 \\
\hline
4 & 0.056 & 0 & 0.9 & 0.146 \\
\hline
5 & 0.043 & 0 & 0.9 & 0.143 \\
\hline
10 & 0.021 & 0 & 0.9 & 0.151 \\
\hline
15 & 0.014 & 0 & 0.9 & 0.153 \\
\hline
20 & 0.010 & 0 & 0.9 & 0.156 \\
\hline
\end{tabular}
\end{table}

\begin{table}
\caption{Description of the decision rule used in the normal phase,
${f}_{norm}$, and the persistence probability-based EB protocol}
\label{tab:ebprot} \centering
\begin{tabular}{|c||c|c|}
\hline
$s_i^{t-1}$ & ${f}_{norm}$ & EB protocol \\
\hline \hline
$idle$ & $p_i^t \approx 1/N$ & \multirow{2}{*}{$p_i^t = p_i^{t-1}$} \\
\cline{1-2}
$busy$ & $p_i^t = 0$ & \\
\hline
$success$ & $p_i^t = 1 - \theta$ & $p_i^t = p_i^{max}$ \\
\hline
$failure$ & $p_i^t \approx 0.5$ & $p_i^t = \max \{ \beta_i p_i^{t-1}, p_i^{min} \}$ ($0 < \beta_i < 1$) \\
\hline
\end{tabular}
\end{table}

\begin{table}
\caption{\textbf{[Protocol 1]} Mission-aware MAC protocol based on
one-period memory} \label{tab:prot1} \centering
\begin{tabular*}{0.93\textwidth}{l}
\hline Decision rule for user $i$ \\
\hline
1. Set $p_i^t = 1$ if $y_i^t = critical$.\\
2. Set $p_i^t = 0$ if $y_i^{t-1} = critical$ and $y_i^t = normal$. \\
3. Set $p_i^t = f_{norm}(s_i^{t-1})$ if
$y_i^{t-1} = normal$ and $y_i^t = normal$.\\
\hline
\end{tabular*}
\begin{flushleft}
$\qquad$ (As in Section 2, we set $y_i^{t'} = normal$ and $s_i^{t'}
= idle$ for $t' \leq 0$ in all protocols.)
\end{flushleft}
\end{table}

\begin{table}
\caption{\textbf{[Protocol 2]} Mission-aware MAC protocol based on
two-period memory} \label{tab:prot2} \centering
\begin{tabular*}{0.93\textwidth}{l}
\hline Decision rule for user $i$ \\
\hline
1. Set $p_i^t = 1$ if $y_i^t = critical$.\\
2. Set $p_i^t = 0$ if $y_i^{t-1} = critical$ and $y_i^t = normal$. \\
3. Set $p_i^t = 0$ if $s_i^{t-2} = success$ and $s_i^{t-1} = failure$. \\
4. Set $p_i^t = f_{norm}(s_i^{t-1})$ if
$y_i^{t-1} = normal$ and $y_i^t = normal$ except for 3.\\
\hline
\end{tabular*}
\end{table}

\begin{table}
\caption{\textbf{[Protocol 3]} Mission-aware MAC protocol based on
$m$-period memory} \label{tab:prot3} \centering
\begin{tabular*}{0.93\textwidth}{l}
\hline Decision rule for user $i$ \\
\hline
1. Set $p_i^t = 1$ if $y_i^t = critical$.\\
2. Set $p_i^t = 0$ if $y_i^{t-1} = critical$ and $y_i^t = normal$. \\
3. Set $p_i^t = 0$ if $s_i^{t-2} = success$ and $s_i^{t-1} = failure$. \\
4. Set $p_i^t = 0$ if $s_i^{t-m} = \cdots = s_i^{t-1} = failure$ and
$y_i^t = normal$.\\
5. Set $p_i^t = f_{norm}(s_i^{t-1})$ if
$y_i^{t-1} = normal$ and $y_i^t = normal$ except for 3 and 4.\\
\hline
\end{tabular*}
\end{table}

\end{document}